\documentclass[final,3p,times,twocolumn]{elsarticle}

\usepackage{graphicx}
\usepackage{bm}
\usepackage{amssymb}
\usepackage{lineno}
\usepackage{setspace}

\biboptions{comma,square}

\journal{Nuclear Instruments and Methods A}

\newcommand{\minerva}{MINER$\nu$A }

\begin{document}

\begin{frontmatter}

\title{
Arachne - A web-based event viewer for \minerva
}

\newcommand{\Rutgers}{Rutgers, The State University of New Jersey, Piscataway, New Jersey 08854, USA}
\newcommand{\Hampton}{Hampton University, Dept. of Physics, Hampton, VA 23668, USA}
\newcommand{\Dortmund}{Institute of Physics, Dortmund University, 44221, Germany }
\newcommand{\Otterbein}{Department of Physics, Otterbein University, 1 South Grove Street, Westerville, OH, 43081 USA}
\newcommand{\JMU}{James Madison University, Harrisonburg, Virginia 22807, USA}
\newcommand{\Florida}{University of Florida, Department of Physics, Gainesville, FL 32611}
\newcommand{\UCIrvine}{Department of Physics and Astronomy, University of California, Irvine, Irvine, California 92697-4575, USA}
\newcommand{\CBPF}{Rua Dr. Xavier Sigaud 150, Urca, Rio de Janeiro, RJ, 22290-180, Brazil}
\newcommand{\PUCP}{}
\newcommand{\INRM}{Institute for Nuclear Research of the Russian Academy of Sciences, 117312 Moscow, Russia}
\newcommand{\Jlab}{Jefferson Lab, 12000 Jefferson Avenue, Newport News, VA 23606, USA}
\newcommand{\Pittsburgh}{Department of Physics and Astronomy, University of Pittsburgh, Pittsburgh, Pennsylvania 15260, USA}
\newcommand{\Guanajuato}{Lascuraín de Retana No. 5, Col. Centro. Guanajuato 36000, Guanajuato. México}
\newcommand{\Athens}{Department of Physics, University of Athens, GR-15771 Athens, Greece}
\newcommand{\Tufts}{Physics Department, Tufts University, Medford, Massachusetts 02155, USA}
\newcommand{\WM}{Department of Physics, College of William \& Mary, Williamsburg, Virginia 23187, USA}
\newcommand{\FNAL}{Fermi National Accelerator Laboratory, Batavia, Illinois 60510, USA}
\newcommand{\Purdue}{Department of Chemistry and Physics, Purdue University Calumet, Hammond, Indiana 46323, USA}
\newcommand{\MCLA}{Massachusetts College of Liberal Arts, 375 Church Street, North Adams, MA 01247}
\newcommand{\UMD}{Department of Physics, University of Minnesota -- Duluth, Duluth, Minnesota 55812, USA}
\newcommand{\Northwestern}{Northwestern University, Evanston, Illinois 60208}
\newcommand{\UNI}{Tupac Amaru Avenue 210, Lima , Peru}
\newcommand{\Rochester}{Rochester, New York 14610 USA}
\newcommand{\Austin}{Department of Physics, University of Texas, 1 University Station, Austin, Texas 78712, USA}
\newcommand{\USM}{Departamento de F\'isica, Universidad T\'ecnica Federico Santa Mar\'ia, Avda. Espa\~na 1680 Casilla 110-V Valpara\'iso, Chile}
\newcommand{\felixOverride}{Lascurain de Retana No. 5, Col. Centro. Guanajuato, Guanajuato 36000. Mexico}
\newcommand{\agagoOverride}{Sección Física, Departamento de Ciencias,  Pontificia Universidad Católica del  Perú,  Apartado 1761, Lima, Perú,   }


\author[Otterbein]     {N.~Tagg \corref{corresponding_author} }
 \author[Otterbein]    {J.~Brangham}
 \author[Rochester]    {J.~Chvojka}
 \author[Otterbein]    {M.~Clairemont}
 \author[Rochester]    {M.~Day}
 \author[Pittsburgh]   {B.~Eberly}
 \author[felixOverride]{J.~Felix}
 \author[Northwestern] {L.~Fields}
 \author[agagoOverride]{A.~M.~Gago}
 \author[UMD]          {R.~Gran}
 \author[FNAL]         {D.~A.~Harris}
 \author[WM]           {M.~Kordosky}
 \author[Rochester]    {H.~Lee}
 \author[USM]          {G.~Maggi}
 \author[MCLA]         {E.~Maher}
 \author[Tufts]        {W.A.~Mann}
 \author[Rochester]    {C.~M.~Marshall}
 \author[Rochester]    {K.S.~McFarland}
 \author[Rochester]    {A.~M.~McGowan}
 \author[Rochester]    {A.~Mislivec}
 \author[Florida]      {J.~Mousseau}
 \author[Florida]      {B.~Osmanov}
 \author[FNAL]         {J.~Osta}
 \author[Pittsburgh]   {V.~Paolone}
 \author[Rochester]    {G.~Perdue}
 \author[Rutgers]      {R.~D.~Ransome}
 \author[Florida]      {H.~Ray}
 \author[Northwestern] {H.~Schellman}
 \author[FNAL]         {D.~W.~Schmitz}
 \author[UCIrvine]     {C.~Simon}
 \author[UNI]          {C.~J.~Solano~Salinas}
 \author[Rutgers]      {B.~G.~Tice}
 \author[WM]           {J.~Walding}
 \author[Hampton]      {T.~Walton}
 \author[Rochester]    {J.~Wolcott}
 \author[WM]           {D.~Zhang}
 \author[UCIrvine]     {B.P.Ziemer}
 \address[Otterbein]{\Otterbein}
 \address[Rochester]{\Rochester}
 \address[Pittsburgh]{\Pittsburgh}
 \address[felixOverride]{\felixOverride}
 \address[Northwestern]{\Northwestern}
 \address[agagoOverride]{\agagoOverride}
 \address[UMD]{\UMD}
 \address[FNAL]{\FNAL}
 \address[WM]{\WM}
 \address[USM]{\USM}
 \address[MCLA]{\MCLA}
 \address[Tufts]{\Tufts}
 \address[Florida]{\Florida}
 \address[Rutgers]{\Rutgers}
 \address[UCIrvine]{\UCIrvine}
 \address[UNI]{\UNI}
 \address[Hampton]{\Hampton}

\author{The \minerva Collaboration}

\cortext[corresponding_author]{Corresponding author}

\date{November 21, 2011}

\begin{abstract} 
  Neutrino interaction events in the \minerva detector are visually represented with a web-based tool called Arachne. Data are retrieved  from a central server via AJAX, and client-side JavaScript draws images into the user's browser window using the draft HTML 5 standard. These technologies allow neutrino interactions to be viewed by anyone with a web browser, allowing for easy hand-scanning of particle interactions. Arachne has been used in \minerva to evaluate neutrino data in a prototype detector, to tune reconstruction algorithms, and for public outreach and education.
\end{abstract}



\end{frontmatter} 


\section{Introduction}\label{sec:intro}

  The \minerva experiment \cite{McFarland:2006pz} is a dedicated neutrino-nucleus cross-section experiment based in the high-intensity Fermilab NuMI neutrino beam, which provides a high-statistics sample of muon neutrinos in the range of 1-20 GeV. \minerva consists of a fine-grained scintillating bar neutrino detector with various nuclear targets. The core of the \minerva detector consists of extruded plastic triangular bars (``strips'') of scintillator and wavelength-shifting readout fibres. Strips are 3.2 by 1.7~cm in cross-section and arranged to form a hexagonal surface 2~m across. Strips are all arranged perpendicular to the neutrino beam, with each plane of strips at one of three orientations in azimuth. A total of approximately  32000 strips are read out by multi-anode photomultipliers tubes (PMTs), with front-end electronics recording timing and pulse-height from each strip. 
  
  A single \minerva record consists of data collected from multiple neutrino interactions that occur during a 10~$\mu$s beam spill. Each record typically has $\sim$6 neutrino-induced interactions.  The offline reconstruction uses timing information to split each record into ``time slices'', with each time slice representing a candidate interaction. Visualization of neutrino interaction events is desirable for many applications: searching for detector pathologies, hand-checking  computer reconstruction codes and event classification, and presenting data to the general public for purposes of outreach.

  \minerva is using a web-based event viewer, named Arachne\footnote{The mythical figure of Minerva turned the mortal Arachne into a spider to punish her impiety, forever cursed to spin webs.}, to do this visualization. Arachne consists of a web page that runs client-side JavaScript code to retrieve and parse event data through extensive use of XML \cite{XML} (eXtensible Markup Language) and AJAX (Asynchronous JavaScript And XML) \cite{AJAX} technologies. A user need simply type a URL into a web browser to see reconstructed data.
  
  Web-based tools have obvious advantages - they require no end-user installation and run on nearly any client platform and operating system. They can be accessed from anywhere in the world with internet service, allowing for easy collaboration and communication.  Because code is deployed only on the central server, they allow for easy maintenance of bug fixes or feature improvements. The code is usually script-based, allowing rapid development cycles for developers.

Purely server-side tools such as CGI scripts have been used in many applications, but are useful only for static queries. Interactive graphical interfaces tend to be unresponsive and slow, due to lag times waiting for the server to build images and HTML. One way this can be improved is through the use of Java applets, an approach that has been used in other high-energy experiments such as the the Kascade viewer \cite{Schieler:2008zz}, as well as the CMS Tracker monitor \cite{Zito:2005xw} and the WIRED generic event viewer \cite{Ballaminut:2000vy}. However, Java applets require plug-ins that can suffer platform dependencies or security problems, and often require large applications to be downloaded to the client before they can work.

Arachne uses an alternative method, using JavaScript running directly in the client's web browser. The client uses AJAX to pull XML data from the server and present it to the screen. This solution has been used in various other experiments for accessing data discovery systems \cite{Aiftimiei:2008zz,Dolgert:2008zza}, offline batch processing job information \cite{Metson:2008zz}, and detector configuration databases \cite{Pedro:2005zza,Roe:2010zza}.  However, these applications are all used for examining experimental meta-data, but not for examining the event data itself as Arachne does.

Ideally, an event display has multiple interactive views. It should allow the user to zoom in on details, select regions of interest based on cuts of time or signal pulse-height, and examine the properties of reconstructed objects with a fast, responsive graphical interface.  Arachne achieves these goals by using the draft HTML 5 standard \cite{html5}, relying in particular on the new \verb|<canvas>| tag. This element allows the JavaScript to draw shapes directly into the user's window, without requiring either pre-rendering on the server or use of plug-ins such as Java applets or Adobe Flash.  Arachne runs in most standard browsers without the need for software installation, allowing it to work not only on any computer platform but also on smartphones and other mobile devices without any special support.\footnote{Google Chrome, Mozilla Firefox, Apple Safari, and Opera all work well, including derivative products. Both Android and Apple mobile devices are also known to work. The notable exception is Microsoft Internet Explorer, which does not work due to its different support of the HTML 5 draft standard.}

In this paper, we describe the architecture of the back-end and front-end software in Section \ref{sec:arch}, the performance of the viewer in Section \ref{sec:performance}, describe some ways the viewer has been used for science and education in Section \ref{sec:applications}, and conclude with our assessment of the success of the software.

\section{Architecture}\label{sec:arch}

  Data from the \minerva electronics is read out by a Data AQuisition system (DAQ). The DAQ saves this ``raw'' data to disk. Raw data is then reconstructed by one or more offline processes based on the GAUDI framework \cite{gaudi} that calibrate the PMT hits, separate neutrino event candidates using timing information, and reconstruct tracks and other topological elements by pattern recognition algorithms.  The result of this offline processing is then stored to disk as a ROOT \cite{root} N-tuple file.  N-tuple files may include only calibrated hits, or may also include mid- and high-level reconstruction objects.  These files form the basic input for Arachne.

\begin{figure}[tb] \begin{center}
   \includegraphics[width=0.95\columnwidth]{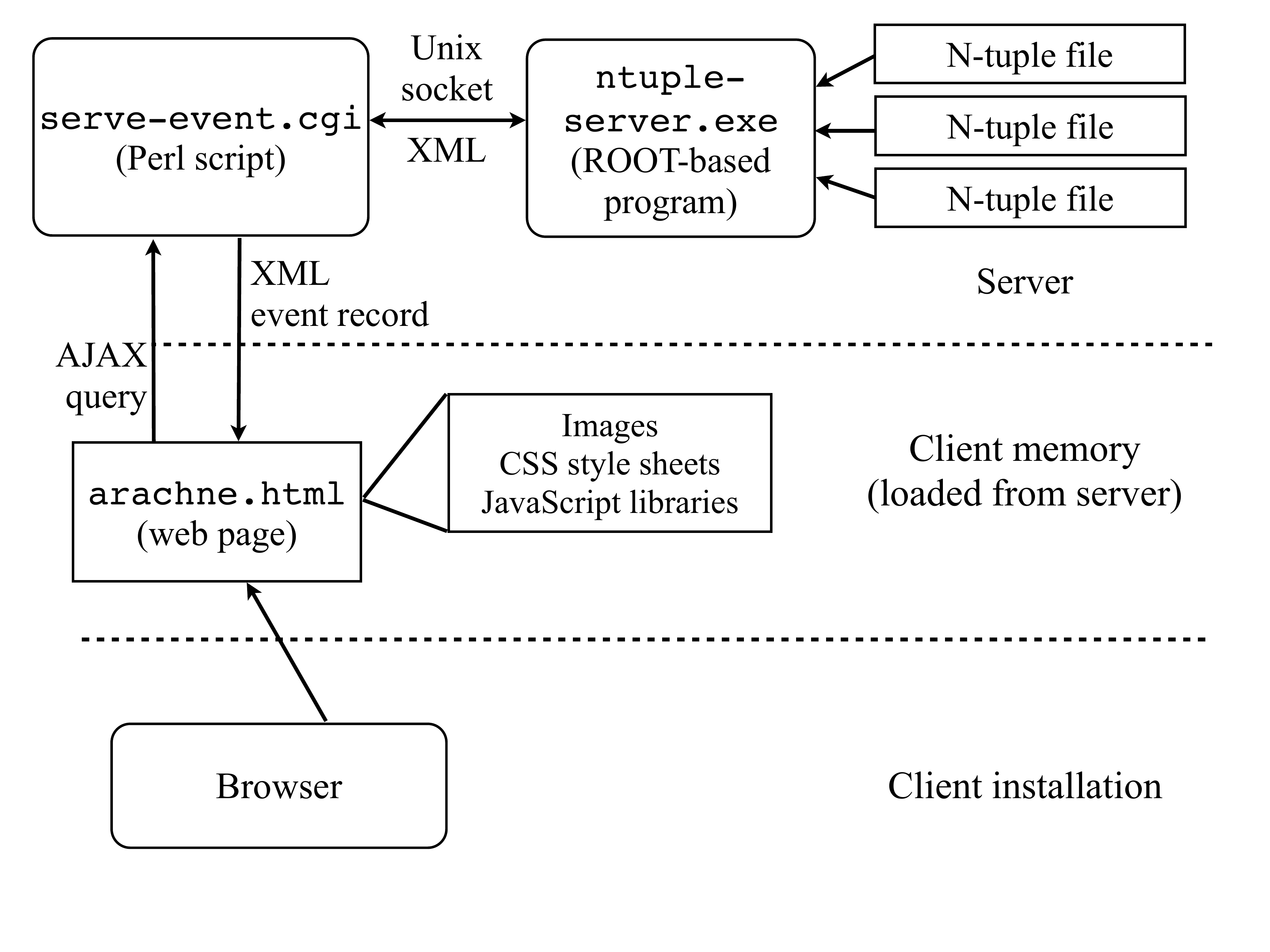}
   \caption{Simplified architecture of the event viewer.}
   \label{f:arch}
   \end{center} \end{figure}

  The basic architecture of the Arachne system is shown in Figure \ref{f:arch}. A user initiates a session by loading the web page \verb|arachne.html| in their browser, which in turn pulls in CSS (Cascading Style Sheet) files, JavaScript libraries, and a few static images, all totaling less than 1 MB.  When the user pushes the ``next event'' button, or otherwise requests new data, a script is launched to initiate an  XHttpRequest \cite{xmlhttprequest} call to the server. This call constitutes a behind-the-scenes HTTP request to a URL on the Arachne server. Data is fetched by the server and returned to the JavaScript for it to draw displays.

\subsection{Server-Side Implementation} \label{sec:server}
   Requests are served by the \verb|serve-event.cgi| Perl script, which finds the file system location of the ROOT N-tuple file containing the requested event.  The location is passed via TPC socket to a static background process called \verb|ntuple-server| along with the record specification. 
   
   The \verb|ntuple-server| is a ROOT-based executable, which opens the N-tuple file and finds the requested record. Entries may be accessed by run and event number, filename, or via user-specified selection criteria. For example, a user could request ``the next two-track event after event 21'', or can request the last event in a live event stream from the detector. The \verb|ntuple-server| then transcribes the N-tuple entry from a row in the ROOT tree into an XML document describing the event. This transcription makes use of the ROOT schema evolution system to parse the N-tuple file, making use of the TStreamInfo classes to interrogate the ROOT file for the needed data. This system allows the \verb|ntuple-server| process to be insensitive to small changes in the N-tuple file format - new variables will not break or confuse the process, and backwards compatibility is easy to maintain. The offline software output schema can evolve without requiring constant updates to the \verb|ntuple-server|.
   
   The \verb|ntuple-server| fills an XML document with the hit and reconstructed object information in a format that is most easily parsed by the client.  This document is then returned to the \verb|serve-event.cgi| script by TPC socket which in turn streams the XML data back to the client browser.  All of this activity on the server has minimal resource load. A typical request for a $\sim$200~kB ROOT record typically results in a $\sim$1~MB XML record that is returned to the client in under one second.

  This server system is agnostic about the data consumer. The data request consists simply of an HTTP GET request, with arguments to specify the search. The returned XML files are self-documenting and can be used by any process, including alternate display packages or online monitoring services. In principle XML data could be generated at any stage of data processing, not just from the N-tuple files. For instance, the DAQ could stream XML records directly to the event display.

\subsection{Client-side Implementation}\label{sec:client}

  When an XML file is successfully delivered to the client, JavaScript scripts are triggered to read the returned event data.  JavaScript built-in tools and third-party libraries such as jQuery \cite{jquery} give powerful and efficient parsing tools for XML data, allowing cuts and slices on elements in the record. An event system is used to trigger each screen component to reload its arrays and re-draw on the screen.   Textual content is rendered by writing HTML into the page, formatted by style sheets.  Graphical data is drawn into \verb|<canvas>| elements using simple primitives such as lines and triangles. 
  
\begin{figure*}[tb] \begin{center}
  \includegraphics[width=\textwidth]{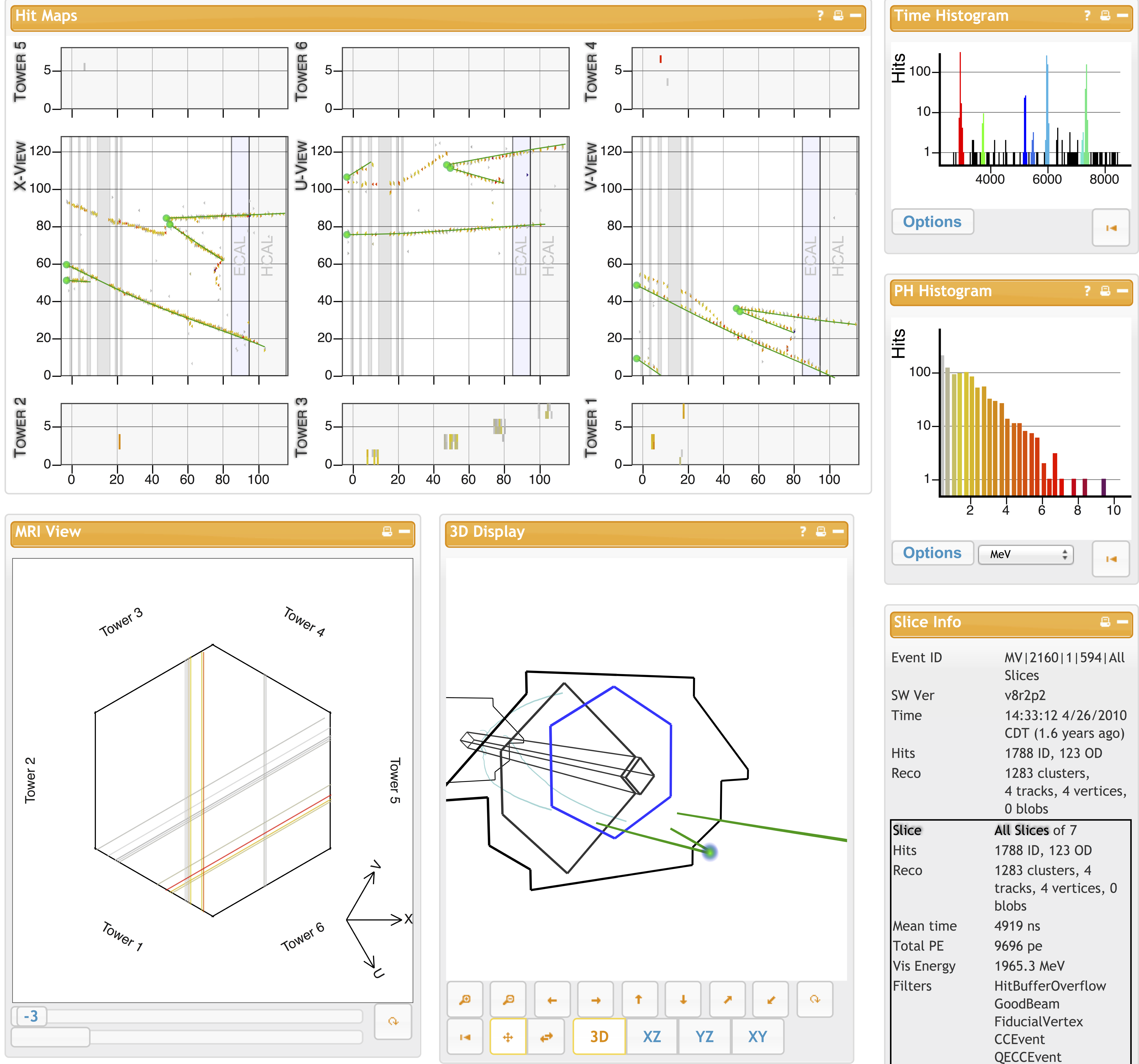}
  \caption{Partial view of the Arachne Event Display.  The right column shows histograms of hit timings (top) and energy (middle), and event statistics (bottom). Top left shows hit maps for the three detector views, along with top and bottom views of the outer calorimeter detector. The lower left shows a longitudinal slice of the hit strips in the detector, which can be scanned along the length of the detector by manipulating the slider control. The bottom middle view is a three-dimensional representation of reconstructed tracks, which can be rotate and panned by the mouse in real time.}
  \label{f:whole_page}
  \end{center} \end{figure*}

  Figure \ref{f:whole_page} shows some of the content seen in the Arachne window. Statistical information is shown as text. Individual PMT hits are drawn in two-dimensional hit-maps corresponding to the scintillator strip orientations. Histograms are created and drawn with color-coding that corresponds to the hit-maps. The histograms can be scaled or clipped by clicking-and-dragging to filter hits in the other views. Hits, tracks, and other reconstructed objects can be selected with the mouse, allowing users to bring up dialog boxes to list object properties. Arachne will highlight related objects to show relationships between the selected object and others.

\begin{figure*}[tb] \begin{center}
  \includegraphics[width=\textwidth]{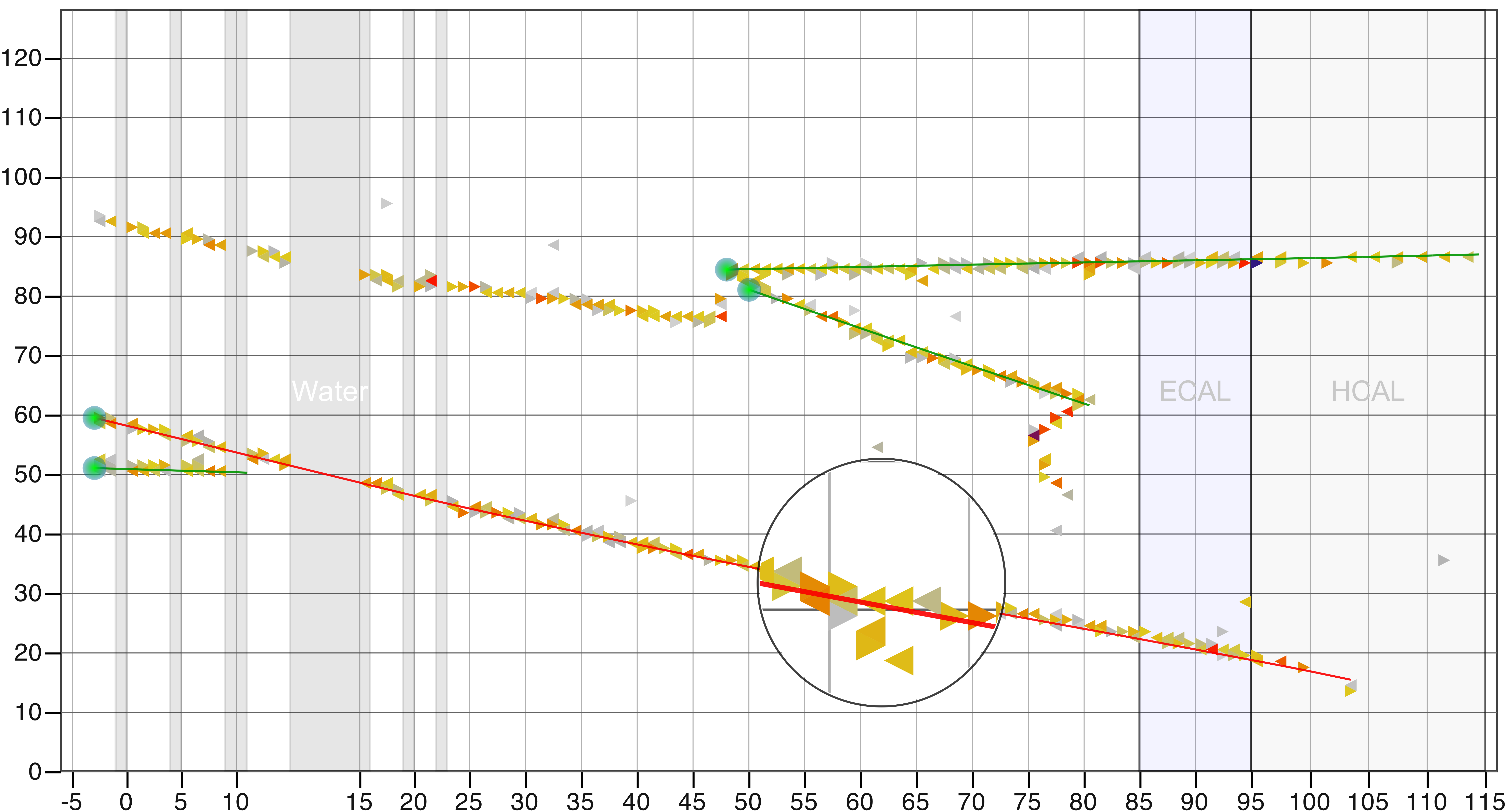}
  \caption{Hit Map. One side-view projection of the detector is shown for a complete 10 $\mu$s beam spill. Colored boxes represent energy depositions in individual scintillator strips, with color denoting magnitude. Overlaid (green) lines show reconstructed particle trajectories. Small circles represent the estimated neutrino interaction points. The circular zoom shows the magnifying-glass effect, which tracks with the motion of the user's mouse in realtime.}
  \label{f:hitmap}
  \end{center} \end{figure*}

  A useful feature of the \verb|canvas| element is that it can provide scaling and clipping. This is employed to create  a magnifying-glass effect around the users' mouse, as shown in Figure \ref{f:hitmap}.  By re-drawing the view with a circular clip region and a scale increase, the user can investigate interesting fine detail without the need to scroll through a large view.
  
  Arachne also provides a 3-D view of the event. The user may click and drag the 3-D image to pan or rotate the view dynamically, with instantaneous re-draws. Objects of interest may be selected by the user to display detailed information.  At present, 3-D rendering is performed using a custom JavaScript library to perform perspective transformations on 2-D primitives to create drawings inside an ordinary \verb|canvas|. This technique is sufficiently fast to draw simple detector wireframes and the $\sim$10 tracks minerva reconstructs per event.  Future versions of Arachne may employ the rapidly-emerging WebGL \cite{webgl} standard to do 3-D rendering, and it may even replace the \verb|canvas| element for 2-D rendering, but this standard is not yet widely adopted.
  
  In sum, the web page is a true application, not just a passive document. User mouse and keyboard events are used to trigger further JavaScript actions. Histograms of pulse-height and timing can be shifted or rescaled dynamically by the mouse, which in turn can create cuts for the hit maps.  Hovering the mouse over a displayed hit will allow hit-specific information to be displayed in hovering dialog boxes. Buttons and selection boxes are used to control both event navigation and display configuration. Through use of jQuery UI enhancements, individual views can be collapsed, moved around the page, and re-sized to provide a user-customized view.  User settings and preferences are saved to browser cookies, so that customizations can be retained between sessions.

\section{Performance}\label{sec:performance}
   Arachne loads data records representing the 10~$\mu$s beam spill. Each \minerva record consists of 2000-6000 hits (after zero-supression), and typically consists of less than 10 reconstructed tracks. The user may display either the entire beam spill, or individual time-slices; performance is roughly the same for both cases. Total time to load and display an event in Arachne is typically between 2 and 10 seconds, depending on the volume of data for the given beam spill.
   
   The XML-encoded files are typically 1-2~MB in size. The data server can locate, encode, and send a single event of this size in under one second. Because data can be sent to the client with http gzip compression, time to send the data by network is short, typically under one second even when accessed at a remote site.  Typical client browsers can unpack and parse the XML document in 0.5 to 1 seconds, and then scan the XML, build histograms, and draw the event to the screen in a time between 1 and 8 seconds.\footnote{Times reported are typical for a 2.5 GHz laptop running Firefox 4.0 or Chrome 11.0. These browsers achieve SunSpider 0.9.1 benchmarks of 350~ms and 318~ms respectively. Older browsers can have longer longer display times due to their older JavaScript excecution engines.}

   Once an event has been built and displayed, the client is responsive. Changing cuts, dragging views, or selecting tracks require redrawing a view in only a few tenths of a second. 


\section{Applications}\label{sec:applications}

\subsection{Comprehensive Hand-scanning}\label{sec:handscan}
   
   Various analyses benefit by a comprehensive hand-scan of an assembly of events. Typically, the organization of such an endeavor between far-flung collaborators can be problematic, but is aided by a tool like Arachne. In the fall of 2009, Arachne was used by a team of hand-scanners to manually classify approximately 18000 events recorded by the \minerva ``tracking prototype'', essentially a quarter-sized \minerva without passive targets.  Arachne made this project possible to perform on a short timescale both because of the way Arachne can be used at a remote site with zero installation and the way new features could be rapidly deployed.
   
   Scanners filled a short form within the Arachne page record event-by-event information,included evaluation of fiducial containment and of activity around the primary vertex, a counts of visual track objects,secondary gamma conversions, neutral particle decays and downstream interactions. Observations concerning other topological details or concerning compatible reaction hypotheses could be entered into a comment field. Typical events from this early data were ~500 hits in size, allowing for quick scanning.  When working steadily, scanners achieved a mean time-per-event of about 30 seconds, with most events taking only about 15 seconds.  The quick turn-around allowed a team of only two dozen volunteer, part-time scanners located all over the US and Brazil to organize and complete a comprehensive data appraisal in a matter of weeks.
   
   Collection of scan records was accomplished by maintaining a database table on the Arachne server, which held lists of events for each scanner's ``inbox''. Arachne was set up to use this inbox to find the next pre-filtered event and present it to the scanner. One part of the Arachne display was a short form the scanner filled out for the event, along with a ``submit'' button. When submitted, Arachne uploaded the scan result to a second database table using another XHttpRequest call. This database was used for early analysis of \minerva data.
    
  \subsection{Outreach and Education}
    
  A version of client Arachne style sheets is available for use with high-school and introductory college level physics courses as a teaching aid.  This version of the Arachne page simplifies the display, emphasizes relevant details, and performs some kinematic calculations for the students using the tracking information sent in the XML files.

  A team of MINERvA physicists and high school teachers have designed two curriculum modules which employ MINERvA data visualized and analyzed through Arachne.  Both take advantage of data from a running particle physics experiment to illustrate concepts already part of the high school curriculum.  The first module analyses muons stopping in MINERvA which subsequently decay to $e^\pm\nu\bar{\nu}$.  Students identify the events from a pre-filtered sample rich in stopping muons and measure the visible electron energy and time of the decay electron relative to the original muon.  Students compare the distribution of event times to the expected exponential behavior, derive a muon half-life, and develop a first hand intuition about the randomness of the decay process.  The energy spectrum of the electrons from muon decay can be used with kinematic arguments to show that there are more than two particles in the final state.  In the second module, students identify quasi-elastic muon neutrino candidate interactions, $\nu n\to\mu^-p$ and calculate the traverse momentum of the target neutron in the initial state from the reconstructed muon and proton in the final state and their knowledge of elastic kinematics.  The surprising result for the students is that the speed of the target neutron is a large fraction of light speed, and that quantum mechanical arguments using the uncertainty principle can be used to illustrate the plausibility of the puzzling result.

 Another version of the page is made available to display ``live'' events coming from reconstruction processes that monitor data coming directly from the detector DAQ. This page is useful both for control room monitoring and as an outreach tool to show visitors a sample of the data being collected by the experiment.  The ``live'' feature will be used in the future to take events from an early stage of nearly real time processing to allow students to perform the studies described above with filtered recent events rather than older data sets.  The high school teachers participating in the design of these modules have emphasized the interest among students in seeing real time data and using this as a point of departure for communication with scientists on the experiment.

\section{Conclusions}

  The principle advantage of a web-based system is the ability for users to connect from any location with nothing more than a browser. A fast learning curve allows new collaborators (such as undergraduate research students) to make meaningful contributions to the experiment without extensive technical skills. 
  
   Developing the event viewer as an end-user application has major advantages. The code needs only to be deployed at a single site, eliminating the need for user installation. Upgrades and bug fixes can be deployed and used as soon as they are developed. Often new features can be deployed and used by end-users in a matter of minutes, and users are always certain they are using the best available version at all times.  Web-based tools can leverage technologies commonly used in commercial contexts. Development of JavaScript applications is very fast, with rapid edit/test/edit development cycles. No tools other than text editors are required for development, unlike alternate web-based technologies such as Flash.  
  
  
  In summary, \minerva has adopted Arachne as its' primary data visualization tool, and has used it successfully for casual data inspection, reconstruction tuning, a comprehensive hand-scan of prototype data, and as an outreach and educational tool.  These applications are made possible by the web-based technologies that continue to gain importance commercially and scientifically as the world-wide-web evolves.

\section{Acknowledgements}
This material is based upon work principally supported by the National Science Foundation under Grant No. 085542. The development of educational resources for \minerva is supported by NSF Grant No. 0619727. The \minerva experiment and NuMI beamline are supported by the Fermi National Accelerator Laboratory, which is operated by the Fermi Research Alliance, LLC, under contract No. DE-AC02-07CH11359, including the \minerva construction project, with the United States Department of Energy. Construction support also was granted by the United States National Science foundation under NSF Award PHY-0619727 and by the University of Rochester.  Support for participating scientists was provided by NASA, NSF and DOE (USA) by CAPES and CNPq (Brazil), by CoNaCyT (Mexico), by CONICYT (Chile), by CONCYTEC, DGI-PUCP and IDI-UNI (Peru), and by Latin American Center for Physics (CLAF). Additional support came from Jeffress Memorial Trust (MK), and Research Corporation (EM).  


\bibliographystyle{elsarticle-num}
\bibliography{arachne}

 \end{document}